**Title: Multiple Imputation Diagnostics when using Electronic Health Record Data in Observational Studies: A Case Study**

Nrupen A. Bhavsar*[1, 3], Lingyu Zhou[2], Samuel I. Berchuck[3], Matthew L. Maciejewski[4], Jerome P. Reiter[2]

[1]Department of Surgery, Duke University School of Medicine, Durham, North Carolina, United States

[2]Department of Statistical Sciences, Duke University, Durham, North Carolina, United States

[3]Department of Biostatistics and Bioinformatics, Duke University School of Medicine, Durham, North Carolina, United States

[4]Department of Population Health Sciences, Duke University School of Medicine, Durham, North Carolina, United States

*Corresponding Author

E-mail: nrupen.bhavsar@duke.edu (NAB)

**Abstract**

**Background**: Missing values in electronic health record (EHR) data pose a significant challenge for epidemiologic research. Data can be missing due to unrecorded information, patient refusal to provide information, and care fragmentation across health systems. Traditional methods for handling missing data, like mean imputation, may introduce bias. Multiple imputation (MI) offers a principled solution by generating multiple plausible values based on statistical models. However, MI requires careful model specification and validation of imputations, ideally using multivariate graphical tools. This paper demonstrates the application of such tools to validate MI in a study of chronic kidney disease, assessing cardiovascular outcomes linked to neighborhood socioeconomic status.

**Methods**: This study used data from Duke University Health System (DUHS) and Lincoln Community Health Center (LCHC). Eligible patients had at least one encounter within DUHS or LCHC and had two estimated glomerular filtration rate (eGFR) values <60 mL/min per 1.73 m$^2$ more than 90 days apart between January 1, 2007 and July 1, 2008. Socioeconomic status was assessed using the Agency for Healthcare Research and Quality (AHRQ) index based on census data. The main outcome was a cardiovascular disease-related hospitalization.

**Results**: Participants were mostly older (mean age 73 years), female (64%), and Black (43%). Participants living in lower neighborhood socioeconomic status (nSES) neighborhoods had higher mean systolic blood pressure (SBP: 140 mmHg) and hemoglobin A1c (HbA1c) levels (7.1%) as compared to participants living in higher nSES neighborhoods.

**Conclusion**: We found that a machine learning based approach, Classification and Regression Trees (CART), was the preferred approach to impute missing data. In addition, the distributions of imputed values of systolic blood pressure (SBP) and hemoglobin A1c (HbA1c) were impacted by whether marginal or conditional values of SBP and HbA1c were imputed. The choice of MI had minimal impact on inference and prediction. Future research may want to extend our results and consider how results may differ when using EHR data from multiple health systems.

**Keywords**: electronic health record (EHR), multiple imputation diagnostics, machine learning, observational study

**Background**

Missing values are a critical challenge when using data from the electronic health record (EHR) for epidemiologic research. Data can be missing from the EHR for a variety of reasons. Some variables may not be assessed at the point of care, such as if a patient is not asked about prior diabetes, a weight scale malfunctioned, or a physician ran out of time during a visit to order a hemoglobin A1c test. In other instances, missing data may arise because it was assessed but not recorded because a patient refuses to disclose smoking history, or a patient answers a provider query, but it is not recorded in the medical record. Another cause of missingness arises due to care fragmentation and non-interoperable medical records, such as when a patient is asked about diabetes status while receiving care in another health system but that information is not shared across all the health systems in which the patient is seen.[1] Procedures to address missing data, such as complete case analysis, missing-indicator method,[2] and mean imputation, can be inefficient or even introduce bias,[3-9] so more principled approaches are needed.

One principled and effective way to address these concerns is to use multiple imputation (MI). In MI, the analyst uses statistical models to generate multiple, plausible values of each individual's missing data, resulting in multiple completed data sets.[5] Subsequently, the analyst estimates the quantities of interest in each completed data set and pools the estimates. In this way, MI can propagate uncertainty due to missing data.[1]

To generate plausible imputations,[10] analysts need to make several decisions about the imputation models, including the functional form, variables to include, accounting for non-linear relationships, and how to impute categorical and non-normal variables.[10-15] In addition, the plausibility of the imputations must be checked based on statistical tools and an understanding of the domain science. Many examples of multiple imputation diagnostics emphasize comparing only univariate distributions of imputed and observed values. While useful, these comparisons may miss key features of the imputations, potentially resulting in scientifically implausible values which can lead to inaccurate inferences. Visualization that can examine multivariate distribution (i.e., multivariate graphical tools), improve upon univariate diagnostics by more robustly checking the validity of imputed distributions. Imputations are conducted conditional on the values of important covariates, leading to more accurate values.

In this paper, we illustrate the benefits of using multivariate graphical tools in combination with the underlying domain science to check the plausibility of imputations and inform imputation model specification. We do this in the context of EHR data drawn from two health systems for a study population with chronic kidney disease. We use the completed data sets to quantify associations between neighborhood socioeconomic status and risk for cardiovascular disease outcomes. We aim to describe diagnostic procedures that can be conducted when imputing missing covariate data. We discuss strategies to deal with issues common in chronic disease epidemiology research, including multiple measurements that are missing together (e.g., multiple lipid and blood pressure measurements) and measurements with distributions that differ markedly for subsets of individuals in the data.

The structure of this article is as follows. In the next section, we introduce the EHR data from the two health systems, then briefly review MI and corresponding diagnostic tools and summarize the process of conducting MI with EHR data. We examine several candidate MI models, using diagnostics and multivariate graphical tools to rule out some models—including some arising from default applications of popular MI software—and arrive at a final specification. We then present results from the completed-data analyses and conclude with general suggestions for MI.

**Materials and Methods**

**Electronic Health Record Data**

*Data Source and Patient Population*

We use data derived from the Duke University Health System (DUHS) and Lincoln Community Health Center (LCHC). The DUHS consists of a larger referral hospital, two community hospitals, and a network of outpatient clinics. The LCHC is a federally qualified health center. We linked patient records from the DUHS to EHRs from the LCHC using common patient identification numbers (i.e., patients in ). Patients were eligible for our study population if they had at least one encounter with the DUHS or LCHC and had two estimated glomerular filtration rate (eGFR) values <60 mL/min per 1.73 m$^2$ more than 90 days apart between January 1, 2007 and July 1, 2008. Patients less than 40 years of age, those with end stage renal disease, prevalent cardiovascular disease, cancer, or HIV/AIDS during the "look-back period" (i.e., January 1, 2007 to July 1, 2008) were excluded. Individuals with outlying values of covariates were excluded from the analysis (total cholesterol <20 or >800; low density lipoprotein (LDL) <10 or >500; , high density lipoprotein (HDL) <15 or >125; creatinine < 0 or > 30; Hemoglobin A1c (HbA1c) < 4% or > 15%; systolic blood pressure (SBP) <40 or < 400; diastolic blood pressure (DBP) <40 or < 400). Patients living outside of Durham County or those with missing address also were excluded.

*Exposure: Neighborhood Socioeconomic Status*

To calculate neighborhood socioeconomic status (nSES), we used the Agency for Healthcare Research and Quality (AHRQ) SES index.[16] The index is a weighted combination of information collected through the American Community Survey (ACS), including the percentage of households that average ≥ 1 persons/room; median value of owner-occupied dwelling, percentage unemployed;

percentage living below poverty level; median household income; percentage ≥ 25 years of age with Bachelor's Degree or more; and percentage ≥ 25 years of age with < 12th grade education. It is scaled to lie between 0-100, with a higher number indicative of higher neighborhood socioeconomic status. Previous studies have used this index to represent a geographic area-based (e.g., census tract, block group) measure of the deprivation experienced according to neighborhood.[16-18] For our study, we use a patient's address at their first encounter from January 1, 2007 to July 1, 2008 to define their census block group federal information processing systems (FIPS) codes and assign a nSES value.

*Covariates*

We obtained from the EHR of DUHS and LCHC patient demographics (e.g., age, sex, race, insurance status) and clinical and comorbid characteristics (e.g., total cholesterol, LDL, HDL, blood pressure, HbA1c, creatinine, hypertension, and diabetes). Hypertension was defined using diagnosis codes, single encounter with elevated blood pressure (SBP>=140, DBP>=90), or medication use. Hypertension categories included no diagnosis or medication use, diagnosis but no medication use, and diagnosis and medication use. No patients lacked a diagnosis for hypertension but were taking anti-hypertensive medication. Diabetes mellitus was defined based on diagnosis codes or single outpatient elevated HbA1c with medication (insulin and non-insulin diabetes medication), diagnosis codes or single outpatient elevated HbA1c with non-insulin medication, and no diabetes diagnosis codes or single outpatient elevated HbA1c. These characteristics were assessed during a "look-back period" from January 1, 2007 to July 1, 2008.

*Outcome: Cardiovascular Disease*

The outcome of interest was cardiovascular disease (CVD)-related admission, defined as a hospitalization for incident myocardial infarction, stroke, or congestive heart failure, based on discharge diagnosis. Patients were censored at their last encounter date or administratively on December 31, 2015, whichever came first.

Data used in this study were initially accessed on December 21, 2021. All analyses were conducted using R v4.1.2 (R Foundation for Statistical Computing, Vienna, Austria). The study was approved by the Duke University School of Medicine Institutional Review Board.

**Brief Review of Multiple Imputation**

*MI by Multivariate Imputation by Chained Equations*

One of the most popular approaches to MI is multivariate imputation by chained equations (MICE). In MICE, we specify the model for each variable conditional on all other variables chosen based on domain expertise, and chain all the variables together in a sequence.[19,20] To illustrate the idea, suppose we have a dataset with sex (i.e., male/female), insurance status (yes/no), and HbA1c (continuous). In this example, sex is fully observed, but insurance status and HbA1c have missing values. To implement MICE, we estimate a regression for binary outcomes (e.g., logistic regression) to predict insurance status from sex and HbA1c, which we then use to impute missing values of insurance status. We then estimate a regression for continuous outcomes (e.g., linear regression) to predict HbA1c from sex and insurance status (including imputed values), which we use to impute missing values of HbA1c. We cycle this process repeatedly (e.g., 20 times) and then pool results using approaches such as Rubin's Rules to create a single completed data set.[21] We repeat the process independently *m* times (e.g., m = 10) to create *m* completed data sets for analysis. Routines for implementing MICE are available in the R, STATA, and SAS (via the package IVEWARE) software packages.

Analysts can choose from a variety of conditional regressions in MICE. A common choice is a standard parametric model like linear and logistic regressions. Alternatively, for continuous variables, analysts can use predictive mean matching (PMM). When using PMM to impute values for some variable *Y*, the analyst fits a linear regression model with *Y* as the outcome and computes the predicted value for each individual in the data set. For each individual with a missing *Y*, the analyst finds the individual with an observed *Y* who has the most similar predicted value. The analyst uses that individual's observed *Y* as the imputation. Another set of approaches adapts machine learning techniques into MI engines, such as classification and regression trees (CART) as implemented in the MICE package in R. [19,22,23] For any variable *Y*, CART recursively partitions units into disjoint subsets based on values of the other variables *X*.[24] This partitioning generates a tree structure, with leaves corresponding to the disjoint subsets of units. To impute a value of a missing *Y*, the analyst finds the leaf that *X* falls in, and randomly samples from the observed *Y* values in that leaf. The CART approach handles interaction effects, nonlinear relations, and complex distributions with minimal tuning needed by the analyst. [22]

*MI Diagnostics*

Most MI diagnostics involve comparing empirical distributions of the imputed values to empirical distributions of the observed values.[10] This can be done by computing summary statistics like means and standard deviations of the observed and imputed values.[25,26] Analysts can also visually compare plots of the observed and imputed values. When the empirical distributions of the observed and imputed values differ markedly, analysts can re-examine the MI model specification to see if it needs improvement.

Differences in empirical distributions between imputed and observed values do not automatically imply poor MI model specifications, depending on expectations informed by domain expertise. For example, values of HbA1c may be missing with high frequency for people who do not have diabetes—and hence typically should have lower values—than those of people who have diabetes. In this case, an accurate imputation model for the missing HbA1c should generate imputed values that are lower than the observed values; a difference in the empirical distributions in the expected direction supports the imputation model for HbA1c.

Multivariate tools can improve on univariate diagnostics by accounting for multivariate relations between different variables. As an example, older individuals may have on average, higher total cholesterol values than younger individuals. Similarly, systolic blood pressure values may differ by whether an individual does or does not have diabetes. Conducting MICE within categories of a covariate and then using visualizations such as single axis scatter plots (i.e., strip plots), takes into account correlations between variables and provides for potentially more accurate imputations.

*MI Inferences*

After diagnostic checks are complete, analysts can apply standard statistical methods on each imputed dataset. Analysts then combine the results of the point and variance estimates in each data set using Rubin's rules for multiple imputation inferences.[27] These can be used, for example, to make confidence intervals for estimands of interest. We refer the reader to any of the many excellent review papers on MI for details of how to implement these statistical methods and combine results.[1,28-34] Whereas these papers provided systematic reviews and broad overviews of MI, processes for using MI in survey data, and brief explanations of MI diagnostics, we provide a primer of how to use MI and review diagnostics when using EHR data for epidemiologic research.

**Results**

***Baseline demographic characteristics of participants***

The study population included 4433 individuals who had a Durham County address, an encounter in the Duke University Health System and/or Lincoln Community Health Center and were diagnosed with chronic kidney disease. The study population had a mean age of 73 years and was 43% Black and 64% female (Table 1). The mean age was similar across tertiles of nSES. A greater proportion of residents in the lowest tertile of nSES were Black (71%) and female (70%) as compared to the middle (Black: 36%; Female: 64%) and highest tertiles of nSES (Black: 21%; Female: 62%).

**Table 1: Baseline characteristics of the study population by tertile of neighborhood SES based on observed data values**

| Characteristic | SES Tertile1 (N = 1483) | SES Tertile2 (N = 1534) | SES Tertile3 (N = 1416) | Total (N = 4433) | Percent of subjects with missing data |
|---|---|---|---|---|---|
| Age (yrs) (n) | 1483 | 1534 | 1416 | 4433 | 0 |
| (Mean, std) | 72 (13) | 73 (12) | 75 (12) | 73 (12) | |
| Female | 1036/1483 (69.9%) | 987/1534 (64.3%) | 872/1416 (61.6%) | 2895/4433 (65.3%) | 0 |
| Race | | | | | 0 |
|   Black | 1052/1483 (70.9%) | 553/1534 (36.0%) | 298/1416 (21.0%) | 1903/4433 (42.9%) | |
|   White | 405/1483 (27.3%) | 944/1534 (61.5%) | 1072/1416 (75.7%) | 2421/4433 (54.6%) | |
|   Other | 26/1483 (1.8%) | 37/1534 (2.4%) | 46/1416 (3.2%) | 109/4433 (2.5%) | |
| Total Cholesterol (n) | 1247 | 1292 | 1164 | 3703 | 17% |
| (Mean, std) | 178 (43) | 177 (42) | 175 (42) | 177 (42) | |
| LDL-C (n) | 1241 | 1281 | 1162 | 3684 | |
| (Mean, std) | 101 (35) | 100 (34) | 96 (33) | 99 (34) | 17% |
| HDL-C (n) | 1238 | 1290 | 1157 | 3685 | 17% |
| (Mean, std) | 51 (16) | 49 (15) | 51 (16) | 50 (16) | |
| Systolic BP (n) | 1107 | 1033 | 986 | 3126 | |
| (Mean, std) | 140 (18) | 137 (17) | 134 (16) | 137 (17) | 20% |
| Diastolic BP (n) | 1107 | 1033 | 986 | 3126 | |
| (Mean, std) | 76 (10) | 75 (10) | 73 (9) | 75 (10) | 20% |
| HbA1C (n) | 918 | 860 | 695 | 2473 | 44% |
| (Mean, std) | 7.1 (1.5) | 6.9 (1.3) | 6.6 (1.0) | 6.9 (1.4) | |
| Creatinine (n) | 1483 | 1534 | 1416 | 4433 | 0 |
| (Mean, std) | 1.6 (0.7) | 1.5 (0.8) | 1.4 (0.6) | 1.5 (0.7) | |
| Hypertension | 1419/1483 (95.7%) | 1431/1534 (93.3%) | 1284/1416 (90.7%) | 4134/4433 (93.3%) | 0 |
| Diabetes | 825/1483 (55.6%) | 728/1534 (47.5%) | 562/1416 (39.7%) | 2115/4433 (47.7%) | 0 |
| **Payor Type** | | | | | |
| Private | 132/1371 (9.6%) | 235/1470 (16.0%) | 260/1391 (18.7%) | 627/4232 (14.8%) | |
| Public | 1170/1371 (85.3%) | 1177/1470 (80.1%) | 1095/1391 (78.7%) | 3442/4232 (81.3%) | |
| Self-Pay | 20/1371 (1.5%) | 12/1470 (0.8%) | 7/1391 (0.5%) | 39/4232 (0.9%) | |
| Unknown | 49/1371 (3.6%) | 46/1470 (3.1%) | 29/1391 (2.1%) | 124/4232 (2.9%) | |
| **Outcomes** | | | | | |
| No. CVD Events | 900 (60.7%) | 886 (57.8%) | 778 (54.9%) | 2564 (57.8%) | |

Abbreviations: SES, socioeconomic status; LDL-c, low density lipoprotein cholesterol, HDL-c; high density lipoprotein cholesterol; BP, blood pressure; HbA1C, hemoglobin A1C.

*Baseline clinical characteristics of the study population based on observed data*

Based on the observed values mean total cholesterol was 188 mg/dL and mean LDL cholesterol was 99 mg/dL. Values were similar across tertile of SES (total cholesterol: 175-178 mg/dL; LDL cholesterol 96-101 mg/dL). The observed mean systolic and diastolic blood pressure were slightly higher among individuals within the lowest tertile of nSES (SBP: 140; DBP: 76) as compared to the highest tertile (SBP: 134; DBP: 73) of nSES. Observed hemoglobin A1c values were higher in the lowest tertile of nSES (7.1%) as compared to the highest tertile of nSES (6.6%). Individuals in the lowest tertile of SES were more likely to use public insurance (85%) as compared to individuals in the highest tertile of SES (79%). The prevalence of hypertension and diabetes was higher among individuals in the lowest tertile of nSES (hypertension: 96%; diabetes: 56%) as compared to the highest tertile of nSES (hypertension: 91%, diabetes: 40%). We note, however, that many variables have missing values, which could make the observed data distributions somewhat unrepresentative of the entire study population and thus motivates the use of MI.

*Extent of missing data*

There was no appreciable missingness for age, Charlson Comorbidity Index, cardiovascular disease, history of diabetes or hypertension, marriage status, medications, race, sex, or insurance status. For variables that were missing, figure 1 shows the extent missing data overall and by tertile of SES. Overall, there is notable missingness for specific variables, including BMI (48%), height (48%), weight (31%), measures of blood pressure (systolic and diastolic blood pressure = 29%), and cholesterol (16-17%). The proportion of data missing was largely lower among individuals living in the first tertile of nSES as compared to individuals living in higher SES neighborhoods (i.e., second and third tertile).

**Figure 1: Percent Missing Data Overall and by Tertile of Neighborhood Socioeconomic Status**

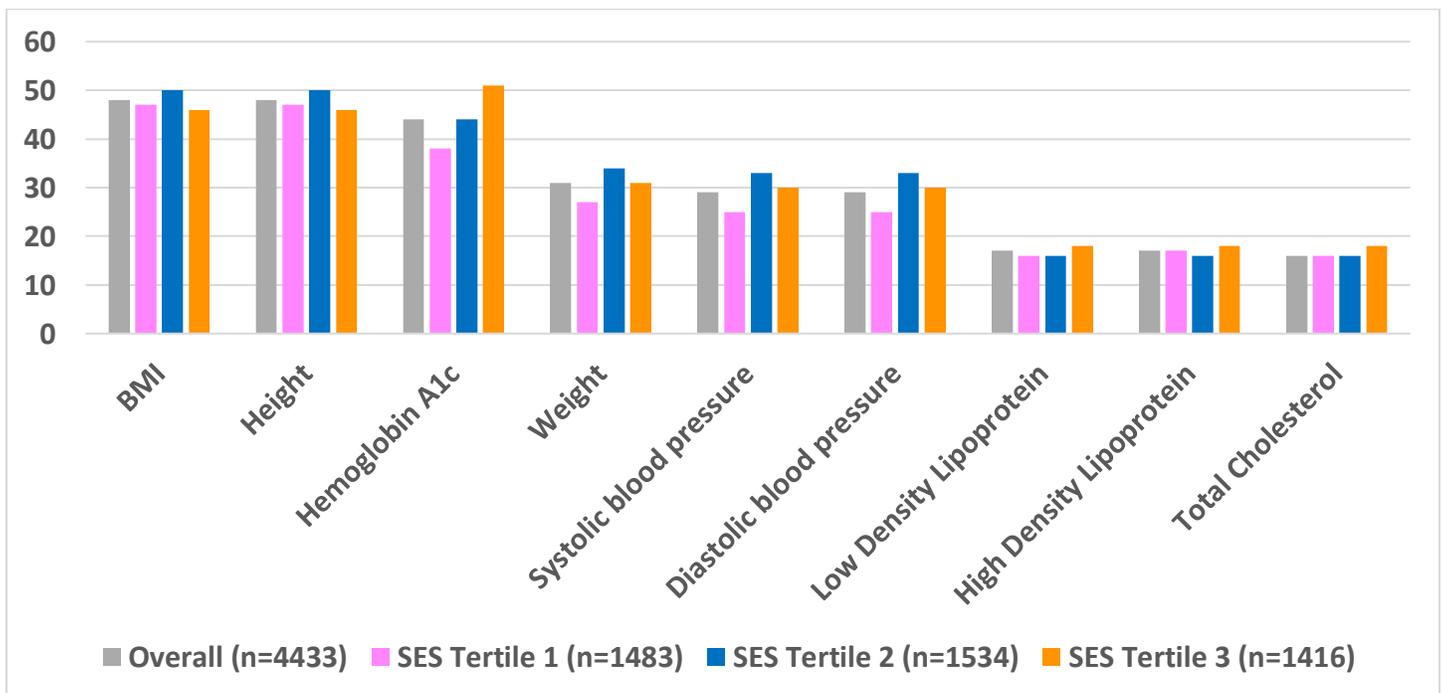

*Specifying the imputation models*

We begin by running MICE under three specifications. The first, which is the default for the software package in R, uses logistic regression for categorical variables (e.g., insurance status, history of diabetes) and PMM for continuous variables (e.g., body mass index (BMI), systolic blood pressure, total cholesterol). The second, termed NORM in the mice package, replaces PMM with linear regression. The third uses Classification and Regression Trees (CART) for all variables. For the parametric models, we adopt a default specification by using main effects only for each variable. To evaluate which specification to use in analyses, we visually compare imputed values with observed values across the three MICE specifications.

Figure 2 displays the empirical marginal distributions of the imputed and observed values for systolic blood pressure (SBP) and hemoglobin A1c (HbA1c). For SBP, the distribution of SBP values is similar across NORM, PMM, and CART, and do not offer any rationale to choose one of over the others. For HbA1c, however, it appears that the imputations for PMM and CART are different from NORM and tend to be concentrated in values between 4-6% while NORM values are well below 4%.

**Figure 2: Marginal distribution of observed and imputed values of systolic blood pressure (SBP) and hemoglobin A1c (A1C) using predictive mean matching (PMM), linear regression (Norm), and Classification and Regression Trees (CART)**

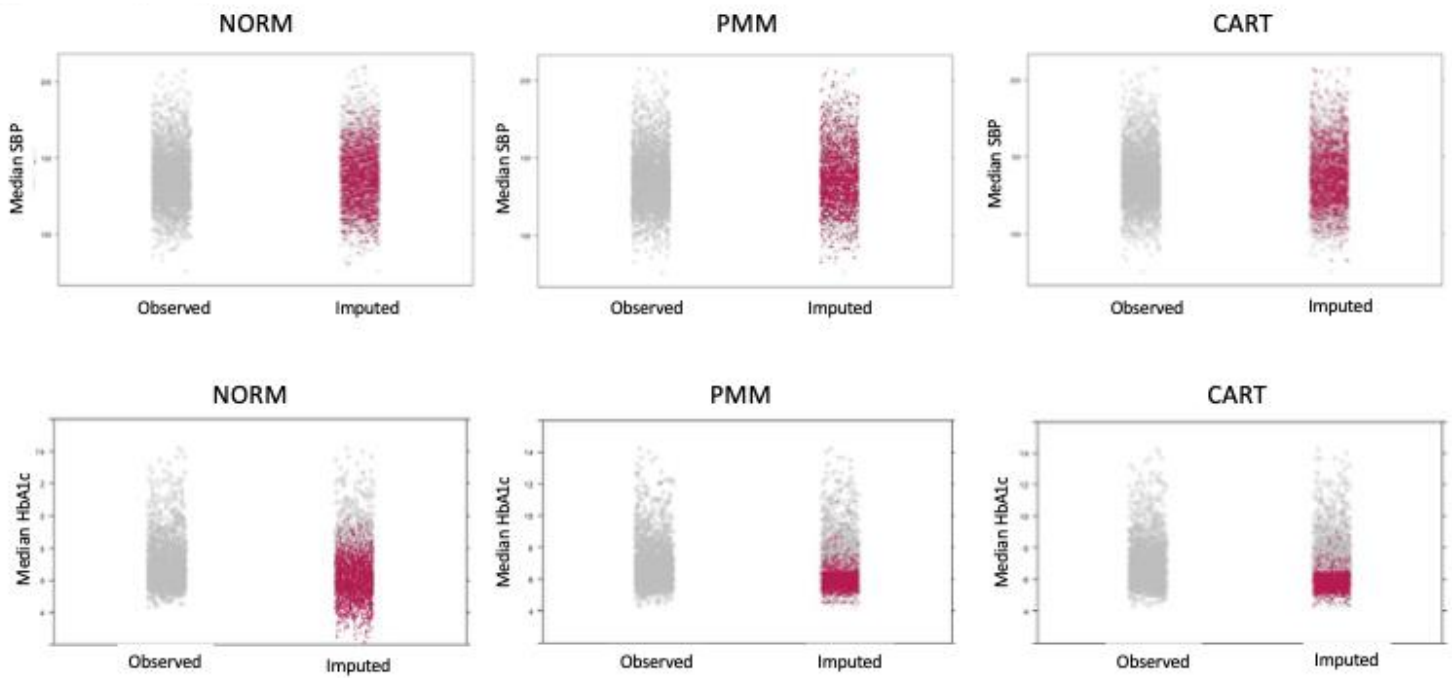

**Gray points = observed**
**Red points = imputed**
**Red points on top of gray points show coverage of imputed points as compared to observed points.**

Further, the imputations for PMM and CART generally are more concentrated at values $\leq 6.0$ compared to the full set of observed values. A naïve interpretation would suggest that the three models, and especially PMM and CART, poorly fit the data. However, an alternative explanation is that values are missing for individuals who do not have diabetes and hence should have lower values of HbA1c. This suggests that the empirical marginal distributions do not provide enough information; we need to evaluate MI diagnostics further. We note that, for the remaining variables, including LDL, HDL, and total cholesterol the empirical distributions of imputed values are reasonably similar across the three methods and do not suggest appreciable differences between the imputed and observed distributions.

Our understanding of the clinical measurements tells us HbA1c values depend on whether or not a person has diabetes, and if these individuals are taking diabetes medication. Similarly, SBP values will vary with the presence or absence of a hypertension diagnosis and medication. Thus, based on this domain knowledge, we should extend diagnostic checks to examine how well the imputations reflect conditional distributions.

Figure 3 displays the empirical distributions of observed and imputed values for HbA1c and SBP conditional on diagnosis and medication status. Specifically, we examine HbA1c values for those with no diagnosis of diabetes or medication use (group 0), no diagnosis of diabetes but taking diabetes medication (group 1), diagnosis of diabetes but not taking diabetes medication (group 2), diagnosis of diabetes and taking diabetes medication (group 3). The diagnostic plots show the poor fit of the PMM and NORM imputations. In particular, when examining HbA1c by diabetes diagnosis and medication status, diagnostic plots for individuals without a prior diagnosis of diabetes and no medication use (i.e., group 0) includes many imputed values exceeding 6.5%, which is clinically implausible and not reflected in the observed data. Fundamentally, the PMM model identifies individuals with observed HbA1c values who have a diabetes diagnosis. The NORM model generates implausible imputations as well; imputed values are out of range for individuals without a diagnosis, medication use, or abnormal laboratory/vital value. Given the relatively poor performance of the PMM and NORM models for HbA1c, CART is the preferred approach for analyses.

**Figure 3: Joint distribution of observed and imputed values of systolic blood pressure (SBP) and Hemoglobin A1c (HbA1c) using predictive mean matching (PMM), linear regression (NORM), and Classification and Regression Trees (CART)**

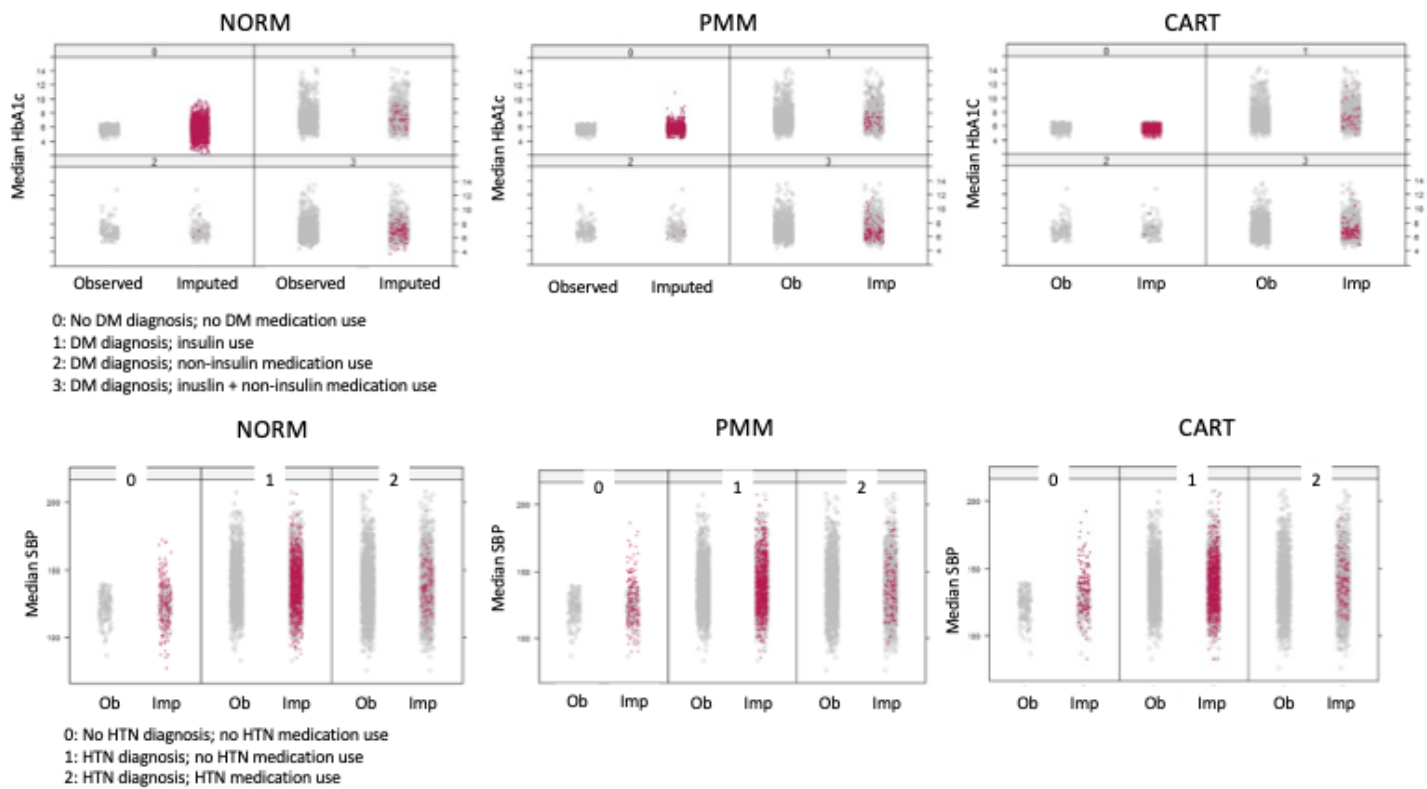

However, the plots for SBP Figure 3 also suggest that we can improve the plausibility of the CART imputations. In particular, we examine SBP values for those with no diagnosis of hypertension or medication use (group 0), no diagnosis of hypertension but taking antihypertensive medication (group 1), diagnosis of hypertension but not taking antihypertensive medication (group 2), diagnosis of hypertension and taking antihypertensive medication (group 3). The CART imputations generate too many high SBP values for individuals in group 0. To correct this, we separate the data by the four groups, resulting in four non-overlapping datasets. We then perform MI using CART models separately in each group. In this way, we allow the CART models to conform specifically to the distributions in each group. Figure 4a displays the resulting imputations of SBP in each group. The partitioning eliminates the implausible high values of SBP in group 0 (individuals without hypertension) without distorting other relationships. We note that the partitioning does not appreciably impact the improvements in HbA1c imputations, as evident in Figure 4b.

**Figure 4a: Observed and imputed values of systolic blood pressure (SBP) (imputed within each SBP group) using Classification and Regression Trees (CART) (grey = observed; red = imputed)**

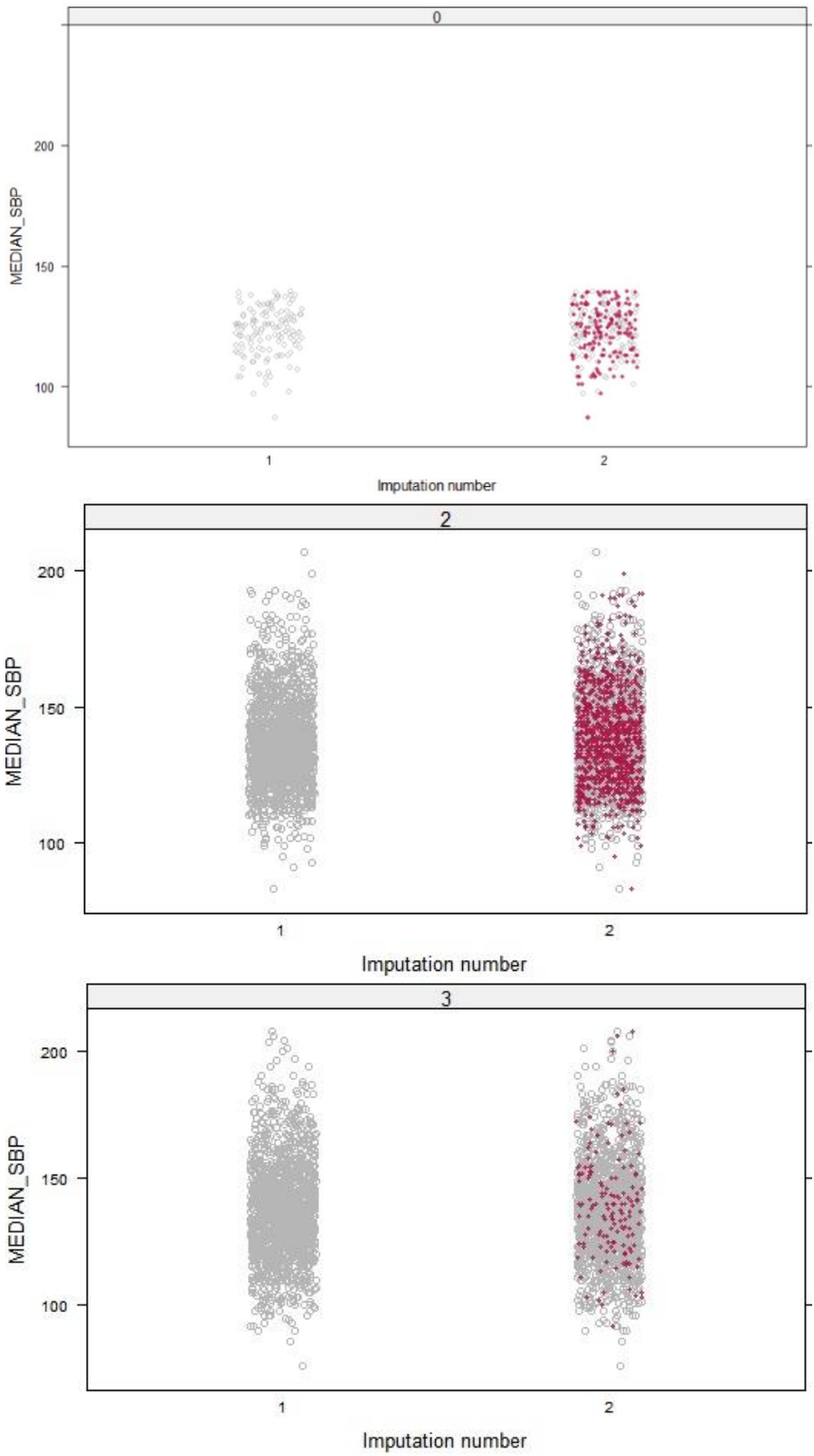

**Figure 4b: Observed and imputed values of hemoglobin A1c (HbA1c) after SBP was imputed within each hypertension group using Classification and Regression Trees (CART) (grey = observed; red = imputed)**

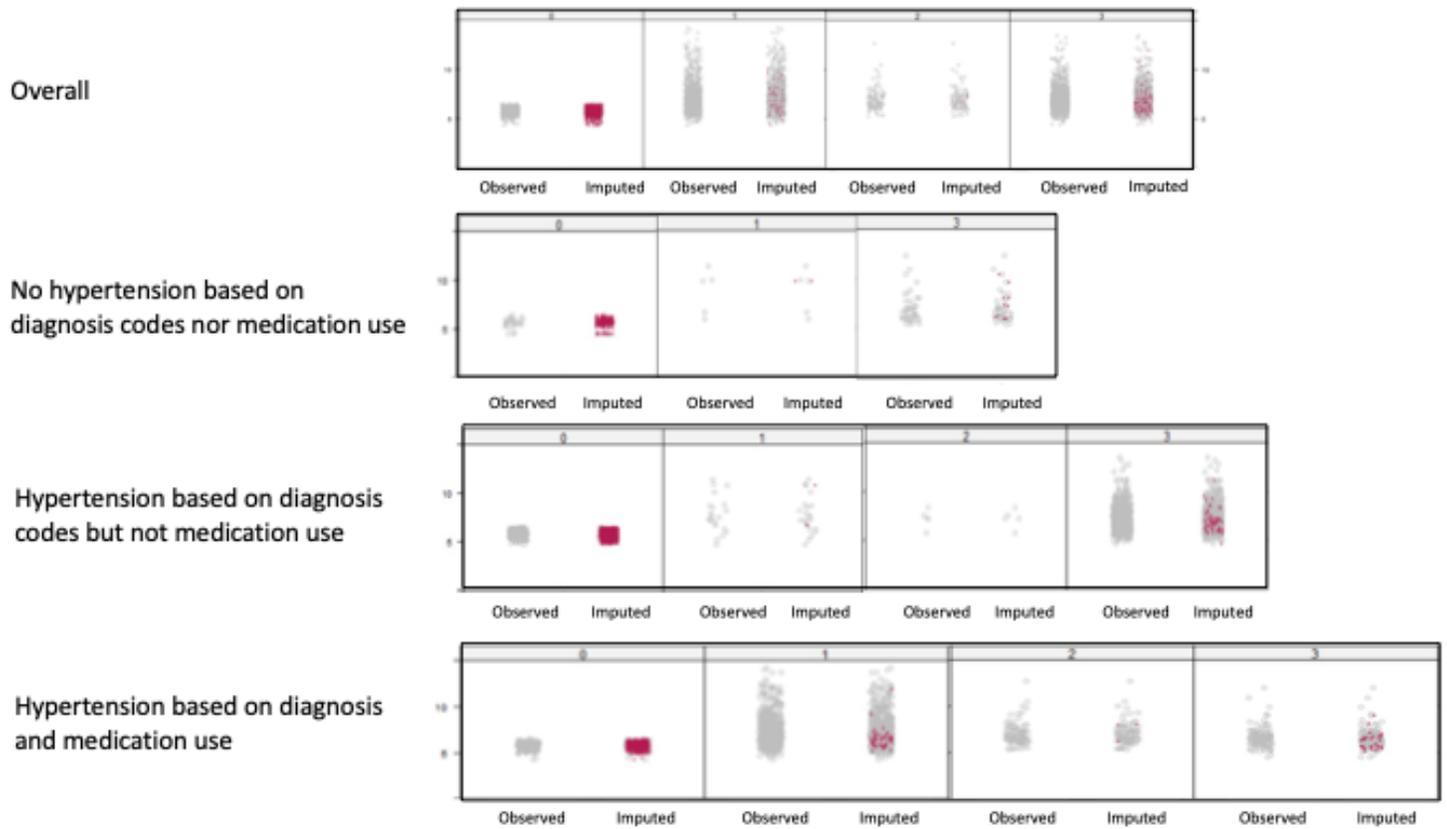

Diabetes Group
0: No DM diagnosis; no DM medication use
1: DM diagnosis; insulin use
2: DM diagnosis; non-insulin medication use
3: DM diagnosis; inuslin + non-insulin medicatio

*Sequential specification for collinear variables*

Before moving to analyses of the completed datasets, we make one comment on a strategy for modeling variables missing together (e.g., lipid measurements [LDL, HDL, and total cholesterol] are either assayed together or derived from each other). When specifying the imputation models in MICE, we first specify a model for LDL conditional on all other variables except HDL and total cholesterol, then specify a model for HDL conditional on LDL and all other variables except total cholesterol, and finally specify a model for total cholesterol conditional on all other variables. This sequential specification, which can be implemented in MICE by specifying which predictors to include in each model, has an advantage over the default treatment in MICE in which one models each variable conditional on all other variables. This default treatment can inaccurately narrow the distribution of imputed values. To illustrate the idea, suppose the correlation between two variables is very high. Then, the default MICE algorithm is likely to generate essentially the same imputed values in each cycle. The sequential specification avoids this problem.

*Analyses after MI*

In complete case analysis with multivariable adjustment for demographic, clinical characteristics, and healthcare utilization, compared to patients in the highest tertile of nSES, patients in the first and second tertiles of nSES had a 11% (HR = 1.11; 95% CI: 0.95, 1.30) and 6% (HR = 1.06; 95% CI: 0.91, 1.25), respectively, greater hazard of a CVD event (Table 2). Interactions between diabetes group and median HbA1c were not statistically significant.

**Table 2: Hazard Ratio (95% Confidence Interval [CI]) for association between neighborhood socioeconomic status (nSES) and hazard for cardiovascular disease (CVD) using complete case and MICE models**

|  | Hazard Ratio (95% CI) |
|---|---|
| **Complete Case** |  |
| nSES Tertile 1 | 1.11 (0.95, 1.30) |
| nSES Tertile 2 | 1.06 (0.91, 1.25) |
| **MICE (CART)** |  |
| nSES Tertile 1 | 1.10 (0.99, 1.22) |
| nSES Tertile 2 | 1.07 (0.96, 1.18) |
| Abbreviations: nSES, neighborhood socioeconomic status; MICE, multivariate imputation by chained equations; CART, Classification and Regression Trees. |  |

Models adjusted for age, race, sex, insurance status, total cholesterol, LDL, HDL, HbA1c, creatinine, diabetes status, hypertension status, number of encounters, Charlson comorbidity index, statin use, and systolic and diastolic blood pressure

In MICE models with CART implementation that used multivariable adjustment for demographic, clinical characteristics, and healthcare utilization, compared to patients in the highest tertile of nSES, patients in the first and second tertiles of nSES had a 10% (HR=1.10; 95% CI: 0.99, 1.22) and 7% (HR=1.07; 95% CI: 0.96, 1.18) respectively, greater hazard of a CVD event. Interactions between diabetes group and median HbA1c were not statistically significant.

*Predicted survival*

We calculated the predicted probability of CVD free survival among individuals who live in low and high nSES (Table 3). The probability of CVD survival was slightly higher between individuals who lived in high, as compared to low, SES neighborhoods. Individuals with higher clinical risk (i.e., prevalent diabetes and hypertension) had lower probability of survival as compared individuals with lower clinical risk (i.e., no prevalent diabetes and hypertension).

**Table 3: Hazard Ratio (95% Confidence Interval [CI]) for Predicted 5-year CVD free survival by neighborhood socioeconomic status (nSES) when imputing using classification and regression trees (CART)**

|  | nSES Category | HR (95% CI) |
|---|---|---|
| **Low Clinical Risk** |  |  |
| Age: 45-64 years<br>Race: White<br>Sex: Male<br>Diabetes: No<br>Hypertension: No | **Low** | 0.69 (0.62, 0.77) |
|  | **High** | 0.72 (0.65, 0.79) |
| Age: 65 years<br>Race: White<br>Sex: Male<br>Diabetes: No<br>Hypertension: No | **Low** | 0.65 (0.58, 0.73) |
|  | **High** | 0.68 (0.61, 0.75) |
| Age: 45-64 years<br>Race: Black<br>Sex: Male<br>Diabetes: No<br>Hypertension: No | **Low** | 0.67 (0.60, 0.76) |
|  | **High** | 0.7 (0.63, 0.78) |
| Age: 65 years<br>Race: Black<br>Sex: Male<br>Diabetes: No<br>Hypertension: No | **Low** | 0.63 (0.56, 0.71) |
|  | **High** | 0.66 (0.59, 0.74) |
| **High Clinical Risk** |  |  |
| Age: 45-64 years<br>Race: White<br>Sex: Male<br>Diabetes: Yes<br>Hypertension: Yes | **Low** | 0.47 (0.39, 0.56) |
|  | **High** | 0.5 (0.43, 0.59) |
| Age: 65 years<br>Race: White | **Low** | 0.41 (0.34, 0.49) |
| Sex: Male<br>Diabetes: Yes<br>Hypertension: Yes | **High** | 0.45 (0.39, 0.52) |
| Age: 45-64 years<br>Race: Black<br>Sex: Male<br>Diabetes: Yes<br>Hypertension: Yes | **Low** | 0.46 (0.39, 0.54) |
| | **High** | 0.49 (0.42, 0.58) |
| Age: 65 years<br>Race: Black<br>Sex: Male<br>Diabetes: Yes<br>Hypertension: Yes | **Low** | 0.4 (0.34, 0.47) |
| | **High** | 0.44 (0.37, 0.51) |
| Abbreviations: nSES, neighborhood socioeconomic status; CI, confidence interval. | | |

**Discussion**

EHR data are unique in that the collection of data does not necessarily follow a regular schedule like traditional epidemiologic studies. As a result, missing data is common and may arise for several reasons. Multiple imputation replaces missing data with plausible values, thereby enabling analysts to account for uncertainty due to missing data.[1] In this paper, we provide an outline of how to conduct and evaluate multiple imputation diagnostics when using data from the EHR. In our example, the distributions of imputed values of SBP and HbA1c were impacted by whether marginal or conditional values of SBP and HbA1c were imputed.

In this context, few studies have discussed missing data, imputation, and diagnostics to guide imputation approaches. Our approach to evaluate multiple imputation diagnostics using EHR data aligns with prior frameworks for addressing missing and conducting diagnostics in non-EHR data. Key steps in this framework include leveraging visualizations and data summaries to evaluate different imputations approaches within MICE, understanding the science to inform imputations that may be unreasonable, and considering conducting imputations within categories of variables to account for how variables vary with one another. Similarly, others have suggested characterizing missingness patterns, compared imputation algorithms, and whether incorporating latent information can improve algorithm performance.[35,36,37]

A limited number of studies have offered guidance on how to conduct imputation diagnostics (broad vs. specific scenarios). Stuart and colleagues provided a discussion of graphical (e.g., histograms, density plots, quantile/quantile plots) and numerical diagnostics (e.g., absolute differences in means between observed and imputed values greater than 2 SDs; risk ratio of variances of observed and imputed values < 0.5 or > 2). Others have provided a more cursory discussion of diagnostics, largely just focusing on model convergence,[31,38,39] and comparing distributions of observed and imputed values. Additional studies have offered approaches under specific situations, such as in the context of Bayesian posterior predictive checks[40] and joint survival models.[40]

**Conclusion**

This primer on MI focuses on observational studies that leverage data from the EHR. We present approaches to evaluate MI diagnostics. Future research can address some of the limitations of the current work. EHR data was used from an urban academic medical center and federally qualified health center that are within the same city. Missing data patterns may differ in areas that are more rural or that contain more health systems. Future research may want to extend the frameworks presented here to these situations. As research increasingly uses EHR data for inferential and prediction studies, missing data will be an important methodological issue to address. The approaches presented in this paper can be used to inform strategies to address missing data.

**List of Abbreviations:**

electronic health record (EHR)

Multiple imputation (MI)

Duke University Health System (DUHS)

Lincoln Community Health Center (LCHC)

estimated glomerular filtration rate (eGFR)

Agency for Healthcare Research Quality (AHRQ)

neighborhood socioeconomic status (nSES)

systolic blood pressure (SBP)

hemoglobin A1c (HbA1c)

Classification and Regression Trees (CART)

low density lipoprotein (LDL)

high density lipoprotein (HDL)

diastolic blood pressure (DBP)

American Community Survey (ACS)

federal information processing systems (FIPS)

cardiovascular disease (CVD)

R (R Foundation for Statistical Computing, Vienna, Austria)

multivariate imputation by chained equations (MICE)

predictive mean matching (PMM)

body mass index (BMI)

linear regression (NORM)

confidence interval (CI)